\documentclass[prl,10pt, twocolumn, nofootinbib]{revtex4-1}

\usepackage{times}

\usepackage{graphicx}
\usepackage{amsmath}
\usepackage[usenames,dvipsnames,svgnames,table]{xcolor}
\usepackage[colorlinks=true,citecolor=blue,linkcolor=black,urlcolor=blue]{hyperref}

\def\ket#1{\left|#1\right\rangle}
\newcommand{\ketbra}[1]{| #1\rangle \langle #1|}
\def\Rb{$^{87}$Rb}
\newcommand{\va}[1]{\ensuremath{(\Delta#1)^2}}
\newcommand{\EQ}[1]{equation~\eqref{#1}}
\newcommand{\vasq}[1]{\ensuremath{[\Delta#1]^2}}
\newcommand{\ex}[1]{\ensuremath{\langle{#1}\rangle}}
\newcommand{\exs}[1]{\ensuremath{\langle{#1}\rangle}}
\newcommand{\EQL}[1]{Equation~\eqref{#1}}

\begin{document} 

\title{Entanglement between two spatially separated atomic modes}
\author{{{Karsten Lange,$^{1}$ Jan Peise,$^{1}$ Bernd L\"u{}cke,$^{1}$ Ilka Kruse,$^{1}$ Giuseppe Vitagliano,$^{2,3}$ Iagoba Apellaniz,$^{3}$ Matthias Kleinmann,$^{3}$ G\'eza T\'o{}th,$^{3,4,5}$ Carsten Klempt$^{1\ast}$}}}
\affiliation{
{$^1$Institut f\"ur Quantenoptik, Leibniz Universit\"at Hannover, Welfengarten~1, D-30167~Hannover, Germany} \\
{$^2$Institute for Quantum Optics and Quantum Information (IQOQI), Austrian Academy of Sciences, Boltzmanngasse 3, A-1090 Vienna, Austria}\\
{$^3$Department of Theoretical Physics, University of the Basque Country UPV/EHU, P.O. Box 644, E-48080 Bilbao, Spain}\\
{$^4$IKERBASQUE, Basque Foundation for Science, E-48013 Bilbao, Spain}\\
{$^5$Wigner Research Centre for Physics, Hungarian Academy of Sciences, P.O. Box 49, H-1525 Budapest, Hungary}\\
{$^\ast$ E-mail: klempt@iqo.uni-hannover.de \:\:\:\:\:\:\:\:\:\:\:\:\:\:\:}
}

\begin{abstract}
Modern quantum technologies in the fields of quantum computing, quantum simulation and quantum metrology require the creation and control of large ensembles of entangled particles.
In ultracold ensembles of neutral atoms, highly entangled states containing thousands of particles have been generated, outnumbering any other physical system by orders of magnitude.
The entanglement generation relies on the fundamental particle-exchange symmetry in ensembles of identical particles, which lacks the standard notion of entanglement between clearly definable subsystems.
Here we present the generation of entanglement between two spatially separated clouds by splitting an ensemble of ultracold identical particles.
Since the clouds can be addressed individually, our experiments open a path to exploit the available entangled states of indistinguishable particles for quantum information applications.
\end{abstract}

\maketitle

The progress towards large ensembles of entangled particles is pursued along two different paths.
In a bottom-up approach, the precise control and characterization of small systems of ions and photons is pushed towards increasingly large system sizes, reaching entangled states of 14 ions~\cite{Monz2011} or 10 photons~\cite{wang2016}.
Complementary, large numbers of up to 3,000 entangled ultracold atoms~\cite{McConnell2015,Haas2014,Hosten2016} can be generated, where the state characterization is advanced top-down towards resolving correlations on the single-particle level.
Because the atoms cannot be addressed individually, ultracold atomic ensembles are controlled by global ensemble parameters, such as the total spin.
Ideally, the atoms are indistinguishable, either with respect to the observable, such as the spin in hot vapor cells~\cite{Julsgaard2001}, or in all quantum numbers in Bose-Einstein condensates (BECs)~\cite{Esteve2008,Gross2010,Riedel2010,Lucke2011,Hamley2012,Berrada2013}.
Is it possible to make these particles distinguishable --- and addressable --- again, while keeping the high level of entanglement?

\begin{figure}[ht!]
	\centering
	\includegraphics[width=0.5\textwidth]{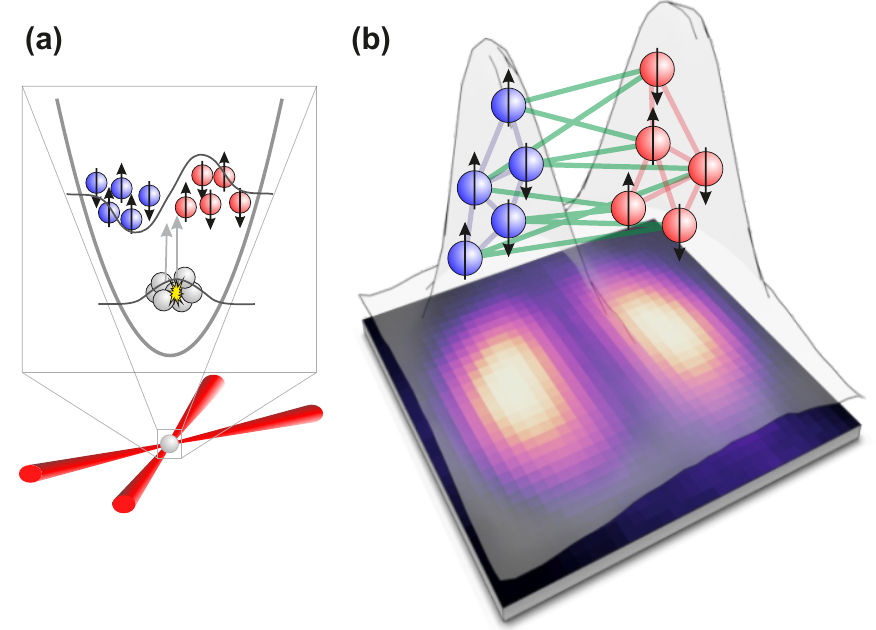}
	\caption{\textbf{Generation of entanglement between two spatially separated atomic clouds. (a)}~A Bose-Einstein condensate of atoms in the Zeeman level $m_F=0$ is prepared in a crossed-beam optical trap.
Collisions generate entangled pairs of atoms in the levels $m_F=\pm 1$ (spin up/down), in the first spatially excited mode.
The created multi-particle entangled ensemble is naturally divided into two clouds (red and blue).
\textbf{(b)} The atomic density profile obtained from an average over 3,329 measurements is shown in the background. The entanglement between the two clouds (indicated by green lines) in the system can be detected by analyzing spin correlations.
		\label{fig1}		
	}
\end{figure}

The generation of entanglement in these systems is deeply connected with the fundamental indistinguishability of the particles~\cite{Killoran2014}.
As an example, two indistinguishable bosons 1 and 2, that are prepared in two independent modes $\rm{a}$ and $\rm{b}$, are described by an entangled triplet state $\frac{1}{\sqrt{2}} (\ket{\rm{a}}_1 \ket{\rm{b}}_2 + \ket{\rm{b}}_1 \ket{\rm{a}}_2)$ due to bosonic symmetrization.
Although this type of entanglement seems to be artificial, the state presents a resource for a Bell measurement~\cite{Yurke1992}.
Equivalently, an ensemble containing the same number of distinguishable spin-up and spin-down atoms is not necessarily entangled, while a twin-Fock state, the corresponding ensemble with
indistinguishable bosons, exhibits full many-particle entanglement~\cite{Lucke2014,Luo2017}.
This form of entanglement is directly applicable for atom interferometry beyond the Standard Quantum Limit~\cite{Lucke2011}.
However, most quantum information tasks require an individual addressing of sub-systems. Despite the experimental progress in entanglement creation in BECs, including the demonstration of Einstein-Podolsky-Rosen correlations~\cite{Peise2015a} and Bell correlations~\cite{Tura2014,Tura2015,Schmied2016}, as well as the demonstration of strongly correlated momentum states~\cite{Bucker2011,Lopes2015a,Dussarrat2017}, a proof of entanglement between spatially separated and individually addressable subsystems has not been realized so far. The possible applications of such a resource reach far beyond quantum information, ranging from spatially resolved quantum metrology to tests for fundamental sources of decoherence or Bell tests of quantum nonlocality.

In this Letter, we report the creation of particle entanglement in an ensemble of up to 5,000 indistinguishable atoms and prove entanglement between two spatially separated clouds. 
We utilize spin dynamics in a spinor BEC to create highly entangled twin-Fock states in a single spatial mode, which naturally splits into two independent parts.
We record strong spin correlations between the resulting two atomic clouds, and derive a criterion to prove their entanglement.
Our results thus demonstrate that the created entanglement of indistinguishable particles can be converted into entanglement of spatially separated clouds, which can be addressed individually.
The concept can be extended to larger numbers of subsystems, down to single particles in an optical lattice, and opens a path to create highly entangled states for numerous applications in quantum information.
For example, it presents a resource to synthesize any pure symmetric state with only single-particle projective measurements~\cite{Wieczorek2009,Kiesel2010}.

\begin{figure}[ht!]
	\centering
	\includegraphics[width=0.5\textwidth]{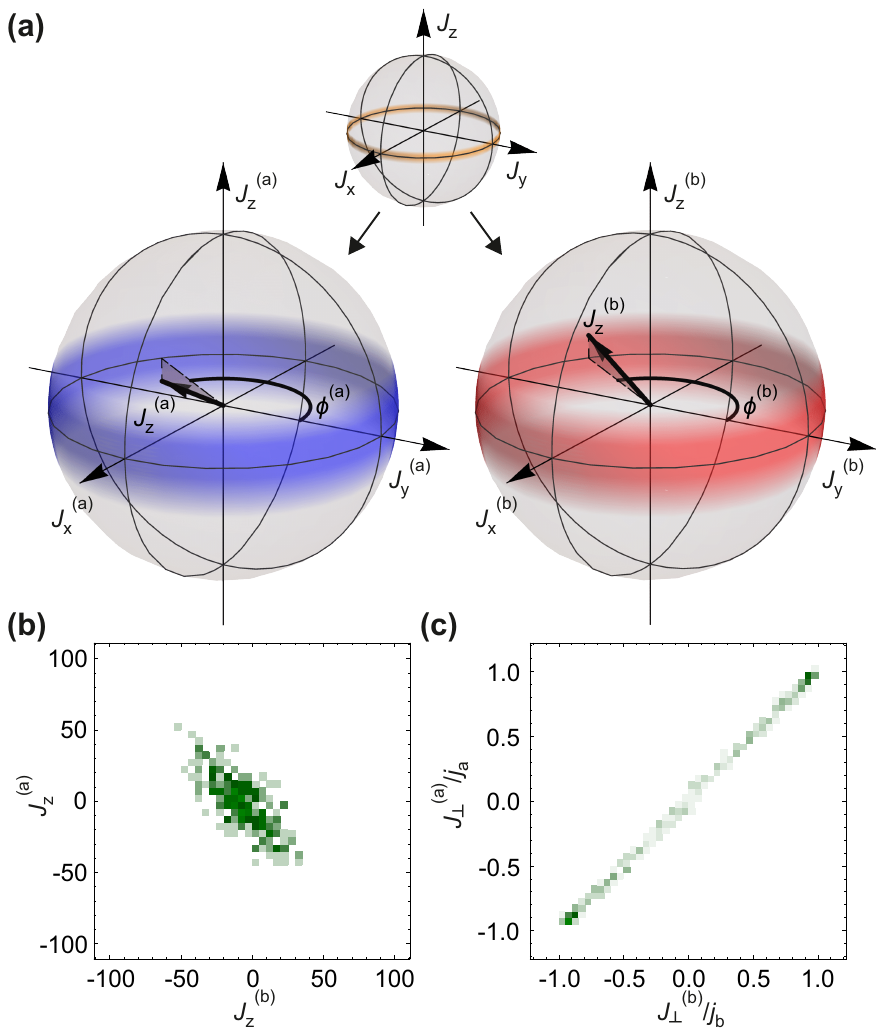}
	\caption{\textbf{Spin correlations between the clouds $\rm{\mathbf{a}}$ and $\rm{\mathbf{b}}$. (a)}
A twin-Fock state is represented by a narrow ring on the equator of the multi-particle Bloch sphere (orange).
When the system is split into two parts $\rm{a}$ (blue) and $\rm{b}$ (red), the states in the single subsystems seem to gain uncertainty.
However, a measurement of $J_{\rm{z}}^{(\rm{b})}$ or $\phi^{(\rm{b})}$ on cloud $\rm{b}$ allows for a prediction of the measurement outcome of cloud $\rm{a}$, such that $J_{\rm{z}}^{(\rm{a})}=-J_{\rm{z}}^{(\rm{b})}$ or $\phi^{(\rm{a})}\approx \phi^{(\rm{b})}$.
\textbf{(b)} Histogram of 506 measurements of $J_{\rm{z}}^{(\rm{a})}$ and $J_{\rm{z}}^{(\rm{b})}$ for a mean total number of 3,460 atoms. 
The data show the anticipated anti-correlation between $J_{\rm{z}}^{(\rm{a})}$ and $J_{\rm{z}}^{(\rm{b})}$. 
\textbf{(c)} 
The strong correlations between the angles $\phi^{(\rm{a})}$ and $\phi^{(\rm{b})}$ are recorded by a measurement of their projection on an arbitrary axis $J_\perp$ in the ${\rm{x}}-{\rm{y}}$ plane.
The histogram of $J_\perp^{(\rm{a})}/j_{\rm{a}}$ and $J_\perp^{(\rm{b})}/j_{\rm{b}}$ (487 measurements) also reflects the projection of a ring onto its diameter.
\label{fig2}
}
\end{figure}

Our experiments start with the preparation of a {\Rb}  Bose-Einstein condensate in a crossed-beam optical dipole trap.
The ensemble of 20,000 particles is transferred to the hyperfine level $F=1, m_F=0$.
Spin-changing collisions create entangled atom pairs in the Zeeman levels $m_F=\pm 1$, where both atoms reside in a spatially excited mode of the dipole trap~\cite{Klempt2009,Scherer2010} (see Fig.~\ref{fig1}).
The output state consists of a superposition of twin-Fock states with an equal number of atoms $N_{\pm 1}$ in the two Zeeman levels $m_F=\pm 1$~\cite{Lucke2011}.
Since the total number of particles is measured during detection, the system is well described by single twin-Fock states with one defined particle number.
Self-similar expansion~\cite{Scherer2010} allows for an imaging of the undistorted but magnified density profiles. 
An inhomogeneous magnetic field separates the atoms to record the atomic densities for each Zeeman level.

\begin{figure*}[ht!]
\begin{center}	
\includegraphics[width=\textwidth]{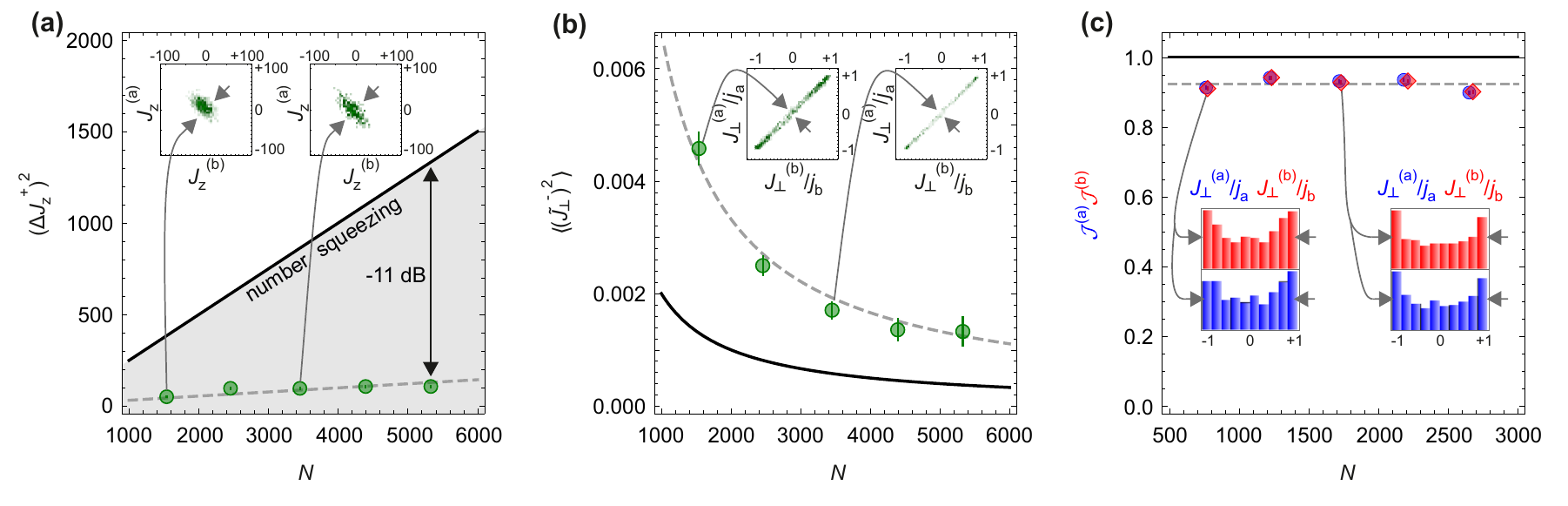}
\caption{\textbf{Spin correlations as a function of the total number of atoms N. (a)}
The prediction variance $(\Delta J_{\rm{z}}^+)^2$ (green circles) surpasses the shot-noise limit (black solid line), indicating number squeezing of up to $-11.0(5)$~dB. The number-dependent detection noise is modeled by a linear fit (gray dashed line).
\textbf{(b)} The fluctuations $\langle (\tilde{J}_\perp^-)^2\rangle =\langle(\tilde{J}_{\rm{x}}^{-})^2\rangle=\langle (\tilde{J}_{\rm{y}}^{-})^2\rangle$ (green circles), corresponding to the phase prediction variance in the experiment, show excess noise, which increases the standard deviation by a factor of $1.8$ (gray dashed line) above the shot-noise limited case (black solid line).
\textbf{(c)}  $\mathcal{J}^{(\rm{a})}$ and $\mathcal{J}^{(\rm{b})}$ quantify the symmetry of the states.
The value 1 for purely symmetric states is indicated in black, the mean experimental value in gray.
In all panels, the total number of atoms $N$ equals the mean of all experimental realizations within an interval of 1,000. The error bars represent one standard deviation of the statistical fluctuations and are obtained via a bootstrapping method (see Supplemental Material).}
\label{fig:crit_separable}
\end{center}
\end{figure*}

The spatially excited mode of the ensembles in $m_F=\pm 1$ provides a natural splitting into a left and right cloud along a line of zero density.
Hence, we divide the initial twin-Fock state into two spatially separated parts $\ket{\rm{a}}$ (left side) and $\ket{\rm{b}}$ (right side).
This process can be described as a beam splitter of the initially populated antisymmetric input mode $\frac{1}{\sqrt{2}}(\ket{\rm{a}}-\ket{\rm{b}})$.
The splitting introduces additional quantum noise due to a coupling with the (empty) symmetric input mode $\frac{1}{\sqrt{2}}(\ket{\rm{a}}+\ket{\rm{b}})$.
In principle, an ideal twin-Fock state shows a maximal entanglement depth~\cite{Lucke2014}, i.e. all atoms that make up a twin-Fock state are entangled with each other.
Therefore, any splitting results in the appearance of quantum correlations between the clouds.

The resulting quantum correlations can be visualized on the multi-particle Bloch sphere (see Fig.~\ref{fig2}(a)).
Here, the atoms in the levels $m_F=\pm 1$ are represented by spin-$\frac{1}{2}$ particles, whose spins $\mathbf{j}^{(k)}$ sum up to a total spin $\mathbf{J}$.
On the Bloch sphere, the lines of latitude represent the number imbalance between the two levels and the lines of longitude represent the phase difference.
An ensemble in a twin-Fock state can be depicted as a ring on the Bloch sphere,
characterized by a vanishing imbalance, $J_{\rm{z}} = (N_{+1}-N_{-1})/2=0$, and an undetermined phase difference.

If we divide the cloud, the collective spins $\mathbf{J}^{(\rm{a})}$, $\mathbf{J}^{(\rm{b})}$ of the two parts
have to sum up to the original collective spin of the full ensemble $\mathbf{J} = \mathbf{J}^{(\rm{a})} + \mathbf{J}^{(\rm{b})}$.
Therefore, the ${\rm{z}}$ components of the collective spins are perfectly anti-correlated $J_{\rm{z}}^{(\rm{a})} + J_{\rm{z}}^{(\rm{b})} = 0$.
Furthermore, since the  spin length $|\mathbf{J}|$ is maximal, the collective spins of the two parts have to point in a similar direction in the $\rm{x}-\rm{y}$ plane and thus have similar azimuthal angles $\phi^{(\rm{a})} \approx \phi^{(\rm{b})}$.

Hence, if the particle number difference of cloud $\rm{b}$ is measured to yield $J_{\rm{z}}^{(\rm{b})}$, the conditioned state of cloud $\rm{a}$ satisfies $J_{\rm{z}}^{(\rm{a})}  = -J_{\rm{z}}^{(\rm{b})}$.
If the value $J_{\rm{x}}^{(\rm{b})}$ ($J_{\rm{y}}^{(\rm{b})}$) is measured on cloud $\rm{b}$, the state of cloud $\rm{a}$ has to fulfill $J_{\rm{x}}^{(\rm{a})} \approx J_{\rm{x}}^{(\rm{b})}$ ($J_{\rm{y}}^{(\rm{a})} \approx J_{\rm{y}}^{(\rm{b})}$).
In summary, the different possible measurements on cloud $\rm{b}$ yield precise predictions for the measurement results of cloud $\rm{a}$, which cannot be explained by a single quantum state that is independent of the chosen type of measurement.
In this sense, the described system is analogous to the thought experiment by Einstein, Podolsky and Rosen~\cite{Einstein1935}, where entanglement is witnessed by the variances of the predictions~\cite{Duan2000a,Simon2000}.
Is it thus possible to detect entanglement between the spatially separated parts of a twin-Fock state?

\begin{figure}[ht!]
\centering
\includegraphics[width=0.5\textwidth]{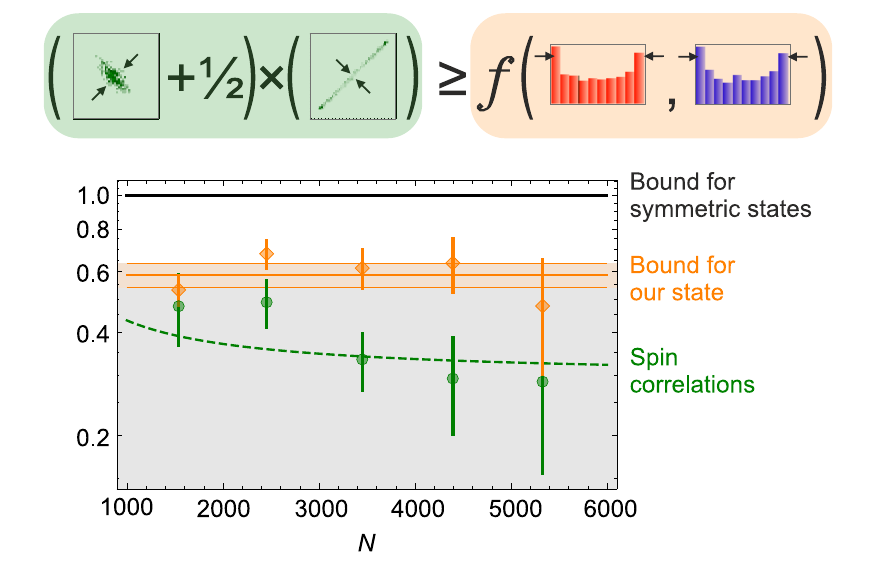}
\caption{\textbf{Violation of the separability criterion as a function of the total number of atoms N.}
The black line represents the right-hand side (RHS) of \EQ{eq:crit_separable} for the ideal case.
The orange line represents the mean  of the experimental results (orange diamonds) of the RHS of \EQ{eq:crit_separable}, where the spin length is reduced.
The green circles show the experimental results for the left-hand side (LHS) of \EQ{eq:crit_separable}.
The dashed green line indicates the prediction of the LHS corresponding to the gray lines in Fig.~\ref{fig:crit_separable}.
The spin correlations clearly violate the criterion by 2.8 standard deviations at a mean total number of 3,460 atoms.
The error bars and shaded orange area indicate one standard deviation and are obtained via a bootstrapping method (see Supplemental Material).}
\label{fig:crit_separable2}

\end{figure}

To this end, we derive an entanglement criterion, which optimally exploits the described spin correlations (see Supplemental Material).
The spin correlations are represented by prediction operators $J_{\rm{z}}^+=J_{\rm{z}}^{  (\rm{a})}+J_{\rm{z}}^{  (\rm{b})},$ and 
$\tilde{J}_{m}^{-}={\tilde{J}}_{m}^{(\rm{a})}-{\tilde{J}}_{m}^{(\rm{b})}$ for $m=\rm{x},\rm{y}$.
Here, the $\rm{x}$ and $\rm{y}$ components are normalized, such that the optimal value is 1, according to 
$\tilde{J}_{m}^{({n})}=\frac{J_{m}^{(n)}/j_n}{\mathcal{J}^{( n)}}$
 with $\mathcal{J}^{(n)}=\left\langle\frac{(J_{\rm{x}}^{(n)})^2+(J_{\rm{y}}^{(n)})^2}{j_n^2}\right\rangle^{\frac 1 2}$ and the spin length $j_n = N_n/2$ for $n={\rm{a}},{\rm{b}}$.
We arrive at a simple separability criterion
\begin{equation}
\left[ \va{J_{\rm{z}}^+}+\tfrac{1}{2}\right] 
\left[\langle (\tilde{J}_{\rm{x}}^{-})^2+(\tilde{J}_{\rm{y}}^{-})^2\rangle\right]\ge f\left(\mathcal{J}^{(\rm{a})},\mathcal{J}^{(\rm{b})}\right),
\label{eq:crit_separable}
\end{equation}
where $f(x,y)=\frac{(x^2+y^2-1)^2}{xy}.$
Any separable state, including mixtures of product states of the form $\sum_k p_k\ketbra{\Psi^{(  \rm{a})}_k}\otimes\ketbra{\Psi^{(  \rm{b})}_k}$ with a fluctuating number of particles, fulfills this inequality. A violation of this criterion indicates that the state is inseparable and therefore entangled.
For perfectly symmetric states, as we would expect in the ideal case, the right-hand side (RHS) of the equation is equal to $1$.
Any deterioration from perfect symmetry is quantified by  $\mathcal{J}^{(\rm{a})}$ and $\mathcal{J}^{(\rm{b})}$.
The inequality has similarities to the famous EPR criterion~\cite{Reid1989} due to the characteristic product of the uncertainties.
It presents a general entanglement criterion, which is particularly sensitive for a spatially separated twin-Fock state.

An application of criterion \eqref{eq:crit_separable} requires an evaluation of the spin correlations between the two clouds $\rm{a}$ and $\rm{b}$.
The measurement results for $J_{\rm{z}}^{(\rm{a})}$ and $J_{\rm{z}}^{(\rm{b})}$ are readily obtained from the absorption images.
The measurement of the orthogonal direction is performed by a sequence of resonant microwave pulses prior to the particle number detection (see Supplemental Material).
The pulses lead to an effective rotation of the spins by $\pi/2$. 
Because the microwave phase is independent of the atomic phases, the rotation yields a measurement of the spin component $J_\perp$ along an arbitrary angle in the $\rm{x}-\rm{y}$ plane.
Since our quantum state is symmetric under rotations around the $\rm{z}$ axis, both due to the initial symmetry and the influence of magnetic field noise,  
the measured distributions of $J_\perp$ can be identified with both $J_{\rm{x}}$ and $J_{\rm{y}}$.
Interestingly, the performed measurement of $J_\perp$ is the
realization of a measurement scheme to demonstrate the violation of a
Bell inequality~\cite{Laloe2009}, if local addressing and a single-particle-resolving
atom counting is added.

Figure~\ref{fig2} shows the histograms of $J_{\rm{z}}$ and $J_\perp/j$ for a mean total number of 3,460 particles in both clouds.
The $J_{\rm{z}}$ data show the expected anti-correlation, while the $J_\perp$ measurements are strongly correlated.
The $J_\perp$~histogram also shows pronounced peaks at the edges, reflecting the projection of a ring onto its diameter.
The strength of these correlations can be quantified by evaluating the prediction uncertainties --- the width of the distributions in the diagonal directions in the histograms, i.e. $\va{J_{\rm{z}}^+}$ and $\langle(\tilde{J}_\perp^{-})^2\rangle$. 

Figure~\ref{fig:crit_separable}(a) presents the prediction variance $\va{J_{\rm{z}}^+}$ as a function of the total number of atoms.
The shown fluctuations, obtained by subtracting independent detection noise, remain well below the shot-noise limit, and are equivalent to a number squeezing of $-11.0(5)$~dB.
The orthogonal quantities (Fig.~\ref{fig:crit_separable}(b)) 
are slightly influenced by technical noise due to small position fluctuations of the clouds,  increasing the standard deviation by a factor of $1.8$ above shot noise. 
Figure~\ref{fig:crit_separable}(c) shows the quantities $\mathcal{J}^{(\rm{a})}$, $\mathcal{J}^{(\rm{b})}$.
We obtain a value of up to $0.94$, close to the ideal value of $1$, indicating a sufficiently clean preparation of an almost symmetric state.

From these results, we can test a violation of the separability criterion.
In Fig.~\ref{fig:crit_separable2}, the orange diamonds correspond to an evaluation of the RHS of the criterion, which would ideally be 1 (black line).
The left-hand side (LHS), represented by the green circles, is well below the RHS, signaling entanglement in the system.
At the best value at a total number of 3,460 atoms, the experimental data violates the separability criterion by 2.8 standard deviations.
Therefore, our measurements cannot result from classical correlations and prove the generation of entanglement between 
spatially separated clouds from particle-entangled, indistinguishable atoms.

Complementary to our work, the group of M. Oberthaler has observed spatially distributed multipartite entanglement and the group of P. Treutlein has observed spatial entanglement patterns.

\section{Acknowledgments}
C.K. thanks M. Cramer for the discussion at the $589.$ Heraeus seminar that led to the initial idea for the experiments, and A. Smerzi for regular discussions and a review of the manuscript. 
This work was supported by the EU (ERC Starting Grant 258647/GEDENTQOPT, CHIST-ERA QUASAR, COST Action CA15220), the Spanish Ministry of Economy, Industry and Competitiveness and the European Regional Development Fund FEDER through Grant No.~FIS2015-67161-P (MINECO/FEDER), the Basque Government (Project No.~IT986-16), the National Research Fund of Hungary OTKA (Contract No.~K83858), the DFG (Forschungsstipendium KL 2726/2-1), the FQXi (Grant No.~FQXi-RFP-1608), and the Austrian Science Fund (FWF) through the START project Y879-N27.
We also acknowledge support from the Deutsche Forschungsgemeinschaft (DFG) through RTG 1729 and
CRC 1227 (DQ-mat), project A02.

\cleardoublepage

\section{Supplemental Material}

\textbf{Experimental sequence and analysis.}
The experimental procedure and evaluation, as well as a discussion of the number-dependent detection noise can be found in detail in the Supplemental Material of Ref.~\cite{Lucke2014}. 

We alternate between the experiments for the two measurement directions $J_{\rm{z}}$ and $J_\perp$ to minimize the influence of changing ambiance conditions. 
Both measurements start with the same experimental sequence. A BEC is prepared in a crossed-beam optical dipole trap in the state $F=1, m_F=0$.

\begin{figure}[ht!]
\centering
\includegraphics[width=0.5\textwidth]{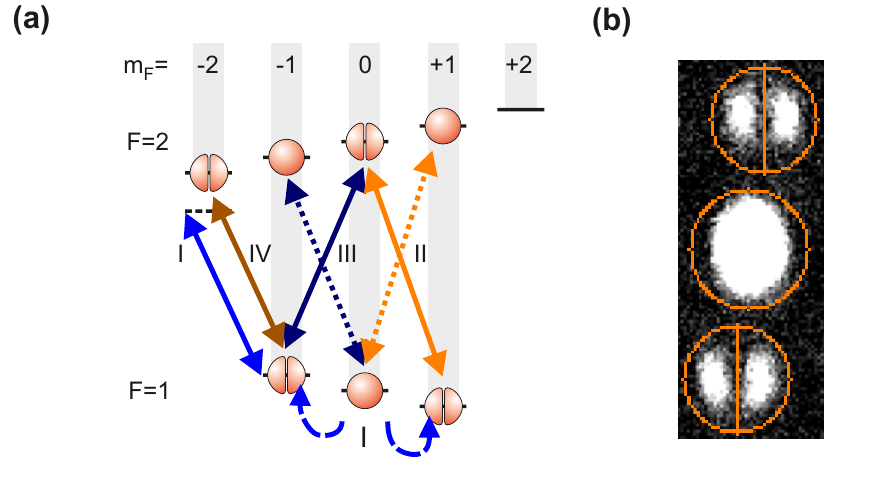}
\caption{\textbf{Experimental techniques. (a)}  Experimental procedure for the measurement of $J_\perp$. The sequence starts with a BEC in the $\ket{F=1, m_F=0}$ Zeeman level .(I) A microwave dressing field (solid blue line) induces a resonance condition for spin-changing collisions, allowing for the population of the $\ket{1, \pm 1}$ levels with a twin-Fock state (dashed blue line). (II) The population in $\ket{1, 1}$ is completely transferred to $\ket{2,0}$ via a microwave transfer (solid orange arrows). This also leads to a population of $\ket{2, 1}$ with atoms from $ \ket{1,0}$ (dotted orange line). (III) A coupling between the two levels populated by the twin-Fock state is induced by a $\pi/2$ microwave pulse (solid dark blue line). This coupling is also resonant to the transition $\ket{1,0} \rightarrow \ket{2,-1}$, populating the  $ \ket{2, -1}$ level (dotted dark blue line). (IV) To avoid an overlap in the detection of the atoms in $\ket{1,-1}$ and $\ket{2, 1}$ the ensemble from  $\ket{1,-1}$ is transferred to  $\ket{2, -2}$ (solid dark orange line). 
\textbf{(b)} Single-shot absorption image with read-out masks indicated in orange. The position of the masks is determined by the center of mass of the central atomic cloud. The other two masks, as well as the cutting lines, are fixed with respect to the central cloud. The atom numbers in all four sub-masks are evaluated by summing over the column densities in the appropriate sub-masks.}
\label{fig:Experiment}
\end{figure}

A red-detuned microwave dressing field with a detuning of $206\,$kHz couples the levels $F=1, m_F=-1$ and $F=2, m_F=-2$. This induces an energy shift of the levels such that a resonance condition for spin-changing collisions from $F=1, m_F=0$ to $F=1, m_F=\pm 1$ is reached~\cite{Lucke2014}. At the resonance, the energy of two atoms in the $m_F=0$ state is equal to the energy of two atoms in $m_F=\pm1$ plus the energy of the excitation to the first spatially excited mode.  This pair creation process, producing a pair of entangled atoms in $m_F=\pm1$, is subject to bosonic enhancement, creating further pairs in the same mode during an interaction time of $t=180\,$ms.
 Due to the nature of the spin-changing collisions, the $F=1, m_F=\pm 1$ levels are populated with a two-mode squeezed state.
The two-mode squeezed state consists of a superposition of twin-Fock states $\sum_n c_n \ket{n}_{+1}\ket{n}_{-1}$ with an equal number of atoms $N_{\pm 1}$ in the two Zeeman levels $m_F=\pm 1$.
The weight $c_n=\frac{(-i \tanh \xi)^n}{\cosh \xi}$ corresponds to a squeezing strength $\xi=\Omega t$ and a spin dynamics rate $\Omega=2 \pi \times 6.6\,$Hz.
The final measurement of the total number of atoms collapses the state onto a twin-Fock state. The measurement of $J_{\rm{z}}$ is now a measurement of the atom numbers in the two levels $m_F=\pm1$ of the $F=1$ manifold. However, to keep the two experimental procedures as similar as possible, the ensembles are transferred to the $F=2$ manifold before detection. To this end, the pulse sequence of the transfer pulses (II-IV) is reversed  for the $J_{\rm{z}}$ measurements  with respect to the $J_\perp$ measurement (see Fig.~\ref{fig:Experiment}).

The measurement in the orthogonal direction requires a rotation around an axis perpendicular to $J_{\rm{z}}$. This is achieved by a coupling of the ensembles in $F=1, m_F=\pm 1$ by an effective $\pi /2$ pulse.
Since the microwave phase is not synchronized to the atomic phases, the rotation leads to a measurement of $J_\perp$ along an arbitrary direction in the $\rm{x}-\rm{y}$ plane. 
However, the state is fully characterized due to its perfect symmetry under rotations around the $\rm{z}$ axis.
Firstly, an ideal twin-Fock state is symmetric itself, and secondly, the experimentally realized state is randomized over $J_{\rm{z}}$ rotations due to the influence of magnetic field noise.
Therefore, the measurement of $J_\perp$ is sufficient. 
 The detection is again realized in the $F=2$ manifold with the $m_F=\pm1$ ensembles occupying $F=2, m_F=-2$ and  $F=2, m_F=0$.  The large condensate from $m_F=0$ is mainly transferred to $F=2, m_F=-1$ with a small fraction transferred to $F=2, m_F=1$.

The experimental sequence ends with the detection of the atomic ensembles. The dipole trap is switched off to allow for $7.5\,$ms of self-similar expansion. The mode profiles remain undistorted but magnified due to the interaction of the ensembles with the large condensate remaining in the $F=2, m_F=-1$ state~\cite{Castin1996}. After the initial mean-field dominated expansion, a strong magnetic field gradient is applied to spatially separate the atoms in the populated Zeeman levels. Finally, the number of atoms in the clouds is detected by absorption imaging on a CCD camera with a large quantum efficiency.

The absorption images are used to detect the number of atoms in the two spatially separated clouds.
The center of mass of the large condensate in the $F=2, m_F=-1$ level is used as a reference for the position of all clouds (see Fig.~\ref{fig:Experiment}(b)). This is necessary due to  slight shot-to-shot variations of the position, which result from minute position changes of the dipole trap. The position of the masks for the ensembles in  $F=2, m_F=\{-2,0\}$ (formerly $F=1, m_F=\pm1$),  as well as the cutting line for the  parts $\rm{a}$ (left) and $\rm{b}$ (right), is fixed with respect to this reference. 
The number of atoms in the four resulting sub-masks is then obtained by summing over the column density of the absorption image to record the spin fluctuations and prove entanglement between the spatially separated atomic clouds.

  To utilize the created state for quantum information tasks, it can be transferred into an optical lattice, where all constituent particles are individually addressable.
As a concrete example, single-atom projective measurements on one half of this highly entangled ensemble allow to synthesize any pure symmetric quantum state in the second half~\cite{Wieczorek2009,Kiesel2010}.

\textbf{Bootstrapping.}
The error bars in Figs.~\ref{fig:crit_separable} and \ref{fig:crit_separable2} are obtained via a bootstrapping method. We created 10,000 random data sets on the basis of the distributions of the experimental data. We then calculated the standard deviations of the measured quantities from these 10,000 samples and checked that the percentage of violations of equation (\ref{eq:crit_separable}) was consistent with the reported significance.

\textbf{Proof of equation (\ref{eq:crit_separable}).}

We start from the sum of two Heisenberg uncertainty relations
$(\Delta J_{\rm{z}})^2[(\Delta J_{\rm{x}})^2 + (\Delta J_{\rm{y}})^2] \ge \frac 1 4 (\ex{J_{\rm{x}}}^2+\ex{J_{\rm{y}}}^2).$
Simple algebra yields
\begin{equation}
	\left[ (\Delta J_{\rm{z}})^2 + \frac 1 4 \right] \times 
		\frac {(\Delta J_{\rm{x}})^2 + (\Delta J_{\rm{y}})^2}
			{\langle J_{\rm{x}}^2 \rangle + \langle J_{\rm{y}}^2 \rangle}
	\ge \frac 1 4 .\label{eq:unc}
\end{equation}
Here, the first factor represents the fluctuations in the particle number difference and the second term represents the fluctuations in the phase difference.

{\it Product states.} First, we consider product states of the form 
$ 
|\Psi^{(  \rm{a})}\rangle\otimes|\Psi^{(  \rm{b})}\rangle. 
$
For such states 
\begin{eqnarray}
&&\left[ \va{J_{\rm{z}}^+}+\frac{1}{2}\right] \times
\left[\va{\tilde{J}_{\rm{x}}^{-}}+\va{\tilde{J}_{\rm{y}}^{-}}\right]
\nonumber\\&&\;\;\;\;\;\;\;\;\;\;=
[{(\mathcal{U}^{({  \rm{a}})}+\tfrac{1}{4})+(\mathcal{U}^{({  \rm{b}})}+\tfrac{1}{4})}]\cdot ({\mathcal{V}^{({  \rm{a}})}+\mathcal{V}^{({  \rm{b}})}}) \nonumber
\\&&\;\;\;\;\;\;\;\;\;\; \ge 4 \sqrt{(\mathcal{U}^{({  \rm{a}})}+\tfrac{1}{4})(\mathcal{U}^{({  \rm{b}})}+\tfrac{1}{4})\mathcal{V}^{({  \rm{a}})}\mathcal{V}^{({  \rm{b}})}}\ge 1
\label{eq:crit_product_proof}
\end{eqnarray}
holds, 
where we used the notation $\mathcal{U}^{({  n})}=\va{J_{\rm{z}}^{(n)}}$ and $\mathcal{V}^{({  n})}=\va{\tilde{J}_{\rm{x}}^{(n)}}+\va{\tilde{J}_{\rm{y}}^{(n)}}$ for $n=\rm{a},\rm{b}$. 
For product states, the variance of a collective observable is the sum of the sub-system variances, i.e. $\vasq{(A^{({  \rm{a}})}+A^{({  \rm{b}})})}=\va{A^{({  \rm{a}})}}+\va{A^{({  \rm{b}})}},$ leading to the equality in \EQ{eq:crit_product_proof}. The first inequality is obtained from the inequality between the arithmetic and the geometric mean.  \EQL{eq:unc} is valid for both part $\rm{a}$ and $\rm{b}$ of the state, leading to the second inequality.

Using
$ 
\exs{(\tilde{J}_{\rm{x}}^{(n)})^2}+\exs{(\tilde{J}_{\rm{y}}^{(n)})^2}=1
$ 
for $n={  \rm{a}},{  \rm{b}},$  \EQ{eq:crit_product_proof} yields
\begin{equation}
2 \left[ \va{J_{\rm{z}}^+}+\frac{1}{2}\right] 
\left(
   \mathcal{S}-\mathcal{C}
\right)
\ge \mathcal{S},
\label{eq:crit_product_simplified_multiplied_by_S}
\end{equation}
where correlations between the two subsystems are characterized by
$ 
\mathcal{C}=\left\langle \frac{J_{\rm{x}}^{(\rm{a})}J_{\rm{x}}^{(\rm{b})}+J_{\rm{y}}^{(\rm{a})}J_{\rm{y}}^{(\rm{b})}}
{j_{  \rm{a}}j_{  \rm{b}}}\right\rangle,
$ 
and
$ 
\mathcal{S}=
\mathcal{J}^{(\rm{a})} 
\mathcal{J}^{(\rm{b})}.
$ 
Note that $\mathcal{C}$ can be negative
and $\vert\mathcal{C}\vert\le \mathcal{S}.$ 
The normalization with the total spin will make it easier to adapt our criterion
to experiments with a varying particle number in the ensembles.

{\it Separable states.} We now consider a mixed separable state of the form
$
\varrho_{\rm sep}=\sum_k p_k \lvert\Psi^{(  \rm{a})}_k\rangle \otimes\lvert \Psi^{(  \rm{b})}_k\rangle.
$ 
 
For such states, we can write the following series of inequalities
\begin{eqnarray}
&&2\left[ \va{J_{\rm{z}}^+}+\frac{1}{2}\right] \left(\mathcal{S}-\mathcal{C}\right)\nonumber\\
&&\;\;\;\;\;\;\;\;\;\;\ge2\left[\sum_k p_k \va{J_{\rm{z}}}_k+\frac{1}{2}\right]\left[\sum_k p_k \left(\mathcal{S}_k-\mathcal{C}_k\right) \right]\nonumber\\
&&\;\;\;\;\;\;\;\;\;\;\ge 2\left[ \sum_k p_k \sqrt{\left(\va{J_{\rm{z}}}_k+\frac{1}{2}\right)\left(\mathcal{S}_k-\mathcal{C}_k\right) } \right]^2\nonumber\\
&&\;\;\;\;\;\;\;\;\;\;\ge \left(\sum_k p_k \sqrt{\mathcal{S}_k}
\right)^2,\label{eq:crit_separable_derivation}
\end{eqnarray}
where the subscript $k$ indicates that the quantity is computed for the $k^{\rm th}$ sub-ensemble $\lvert \Psi^{(  \rm{a})}_k\rangle \otimes\lvert\Psi^{(  \rm{b})}_k\rangle.$
The first inequality in \EQ{eq:crit_separable_derivation} is due to 
$\va{J_{\rm{z}}^+}$ and $\mathcal{S}$ being concave in the quantum state. 
The second inequality is based on the Cauchy-Schwarz inequality
$
\left(\sum_k p_k a_k\right)\left(\sum_k p_k b_k\right) \ge \left(\sum_k p_k \sqrt{a_k b_k}\right)^2,
$ 
where $a_k,b_k\ge 0.$ 
The third inequality is the application of \EQ{eq:crit_product_simplified_multiplied_by_S} for all sub-ensembles. Next, we find a lower bound on the RHS of \EQ{eq:crit_separable_derivation} based on the knowledge of $\mathcal{J}^{{(  \rm{a}})}$ and $\mathcal{J}^{{(  \rm{b}})}.$

We find that
\begin{equation}
\sum_k p_k \left(\mathcal{J}^{(\rm{a})}_k\mathcal{J}^{{(  \rm{b}})}_k\right)^{1/2} 
\ge (\mathcal{J}^{(\rm{a})})^2+(\mathcal{J}^{(\rm{b})})^2-1,
\label{eq:crit_separable_derivation_minimization}
\end{equation} 
which is based on noting $(xy)^{1/4}\ge x+y-1$ for $0\le x,y\le 1.$
Using \EQ{eq:crit_separable_derivation_minimization} to bound the RHS of \EQ{eq:crit_separable_derivation} from below and dividing by $\mathcal{S}$ we obtain
\begin{equation}
\left[ \va{J_{\rm{z}}^+}+\frac{1}{2}\right]\times  \left[2-2\frac{\mathcal{C}}{\mathcal{S}}\right]\ge \frac{\left[(\mathcal{J}^{({  \rm{a}})})^2+(\mathcal{J}^{({  \rm{b}})})^2-1\right]^2}{\mathcal{S}}.
\label{eq:crit_separable_appendix}
\end{equation}

{\it Non-zero particle number variance.} So far, we assumed that the particle number of the two clouds $\rm{a}$ and $\rm{b}$ are known constants. In practice, the particle number is not a constant, but varies from experiment to experiment. In principle, one could postselect experiments for a given particle number, and test entanglement only in the selected experiments. However, this leads to discarding most experiments, increasing the number of repetitions needed tremendously. Hence, we  modify our condition to handle non-zero particle number variances~\cite{Hyllus2012}. In this case, the state of the system can be written as
$ 
\varrho=\sum_{j_{  \rm{a}},j_{  \rm{b}}} Q_{j_{  \rm{a}},j_{  \rm{b}}} 
\varrho_{j_{  \rm{a}},j_{  \rm{b}}},
$ 
where $\varrho_{j_{  \rm{a}},j_{  \rm{b}}}$ are states with $2j_{\rm{a}}$ and $2j_{\rm{b}}$ particles in the two clouds, $Q_{j_{  \rm{a}},j_{  \rm{b}}}\ge 0,$ $\sum_{j_{  \rm{a}},j_{  \rm{b}}} Q_{j_{  \rm{a}},j_{  \rm{b}}}=1.$ 
The state $\varrho$ is separable if and only if all $\varrho_N$ are separable.
Then, expectation values for $\varrho$ are computed as
$ 
\exs{Af(\hat{j}_{  \rm{a}},\hat{j}_{  \rm{b}})}_{\varrho}=\sum_{j_{  \rm{a}},j_{  \rm{b}}} Q_{j_{  \rm{a}},j_{  \rm{b}}} \ex{A}_{\varrho_{j_{  \rm{a}},j_{  \rm{b}}}} f(j_{  \rm{a}},j_{  \rm{b}}),
$ 
where the operator is separated into one part that depends only on the particle number operators of the two clouds represented by $\hat{j}_{  \rm{a}}$ and $\hat{j}_{  \rm{b}}$, and another part that does not depend on them. $f(x)$ denotes some function. The proof from \EQ{eq:crit_product_proof} to \EQ{eq:crit_separable_appendix}
can then be repeated, assuming that $j_{  \rm{a}}$ and $j_{  \rm{b}}$ are operators.
Hence, we arrive at the criterion that can be used for the case of varying particle numbers given in equation (\ref{eq:crit_separable}).

Note that we choose the normalization of the variances such that the criterion is robust against fluctuations of the total number of particles.
For a constant particle number
one could simplify the fractions on the LHS of equation (\ref{eq:crit_separable}) by multiplying both the denominator and the numerator by $j_{  \rm{a}},$ and for the other fractions by $j_{  \rm{b}}.$

\bibliographystyle{apsrev4-1}
\bibliography{Lange}

\end{document}